\documentclass[referee,a4paper,12pt,traditabstract]{swsc} 
\usepackage{graphicx}

\usepackage{tabularx}
\usepackage{adjustbox}

\usepackage{txfonts}
\usepackage{subfigure}
\usepackage{epstopdf}
\usepackage[authoryear,round]{natbib}
\usepackage[backref]{hyperref}
\usepackage{url}
\usepackage{enumitem}
\usepackage{multirow}
\usepackage[table]{xcolor}

\usepackage{color, colortbl}
\definecolor{Gray}{gray}{0.9}

\hypersetup{colorlinks=true,citecolor=cyan,urlcolor=cyan,linkcolor=blue}

\begin{document}

   \title{Investigation of pre-flare dynamics using the weighted horizontal magnetic gradient method:}

   \subtitle{From small to major flare classes}
   
   \titlerunning{ $WG_{M}$ method}

   \authorrunning{M. B. Kors\'os,  S. Yang and R. Erd\'elyi} 

   \author{M. B. Kors\'os \inst{1,2,3}  \and S. Yang \inst{4,5} \and R. Erd\'elyi\inst{1,3} }

   \institute{1. Solar Physics \& Space Plasma Research Center (SP2RC), School of Mathematics and Statistics, University of Sheffield, Hicks Building, Hounsfield Road, S3 7RH, UK\\2. Debrecen Heliophysical Observatory (DHO), Research Centre for Astronomy and Earth Sciences, Hungarian Academy of Science, 4010 Debrecen, P.O. Box 30, Hungary\\ 3. Department of Astronomy, E\"otv\"os Lor\'and University, P\'azm\'any P. s\'et\'any 1/A, Budapest, H-1117, Hungary\\4. CAS Key Laboratory of Solar Activity, National Astronomical Observatories, Chinese Academy of Sciences, Beijing 100012, China\\5. School of Astronomy and Space Science, University of Chinese Academy of Sciences, Beijing 100049, China\\Corresponding email: korsos.marianna@csfk.mta.hu} 
            
\abstract{ There is a wide range of eruptions in the solar atmosphere which contribute to space weather, including the major explosions of radiation known as flares. To examine pre-event behavior in $\delta$-spot regions, we use here a method based on the weighted horizontal gradient of magnetic field ($WG_{M}$), defined between opposite polarity umbrae at the polarity inversion line of active regions (ARs) as measured using from the Debrecen Heliophysical Observatory catalogues. In this work, we extend the previous analysis of high-energy flares to include both medium (M) and low-energy (C and B) flares. First, we found a logarithmic relationship between the log value of highest flare class intensity (from B- to X-class) in a $\delta$-spot AR and the maximum value of $WG_{M}$ of the 127 ARs investigated. We confirm a trend in the convergence-divergence phase of the barycenters of opposite polarities in the vicinity of the polarity inversion line. The extended sample (i) affirms the linear connection between the durations of the convergence-divergence phases of barycenters of opposite polarities in $\delta$-spot regions up to flare occurrence and (ii) provides a geometric constraint for the location of flare emission around the polarity inversion line. The method provides a tool to possibly estimate the likelihood of a subsequent flare of the same or larger energy.} 
  
  
   \maketitle
%%
%%________________________________________________________________

\section{Introduction} \label{Introduction}

A clear understanding of the dynamics and energetics of magnetic reconnection remains an important goal of solar flare research. A key aspect of this research area is to determine the dynamics of an active region (AR) before flare occurrence. Reliably identifying flare precursors also has practical significance for flare predictions \citep[see, e.g.,][and references therein]{Georgoulis2012,Georgoulis2013, Barnes2016}. The aim of finding such precursors was achieved by, e.g., \cite{Korsos2014} and \cite{Korsos2015} (K14 and K15 thereafter, respectively), developing a new type of proxy measure of magnetic non-potentiality in an AR.

K14 and K15 analysed the horizontal gradient of the line of sight (LOS) component of the magnetic field in the vicinity of polarity inversion lines (PILs), and found indicative features of the imminent flaring behaviour up to two-three days prior to the actual flare occurrance. The pre-flare dynamics and the related physical processes at the solar surface were investigated using data with an hourly temporal resolution from joint SDD ground- (Debrecen Heliophysical Observatory, DHO) and space-based (Solar and Heliospheric Observatory, SOHO) sunspot data catalogues \cite[see, e.g.][]{Baranyi2016} from 1996 to 2010. Furthermore, K14 and K15 focused solely on the largest intensity flare-class during the investigated AR's disk passage. Based on the Geostationary Operational Environmental Satellite (GOES) flare classification system, the 61 investigated flare cases were all stronger than M5 flare-class in those two papers.

K14 introduced the horizontal gradient of the magnetic field ($G_{M}$), a proxy which measures the magnetic non-potentiality at the photosphere. The $G_{M}$ is applied between umbrae with opposite polarities at the PIL of ARs and provides key information about the most important properties of an imminent flare: its intensity and occurrence time. K14 identified pre-flare patterns of this proxy quantity: increasing phase, high maximum and gradual decrease prior to flaring. A linear relationship was found between the pre-flare $G_M$ maximum and the largest flare intensity class of the AR. Next, the occurrence time estimation was found to be somewhat less precise, for the most probable time of flare occurrence being between 2-10 hrs after the $G_{M}$ maximum. 

In K15, the authors generalised the $G_{M}$ method and presented the concept of the weighted horizontal gradient of the magnetic field, $WG_{M}$. 
The introduction of the $WG_{M}$ has enhanced the application capability by indicating a second flare precursor. Namely, the barycenters (the area-weighted centers of the positive and negative polarities) display a pattern of approach and recession prior to the flare.
We found that the flare occurs when the distance between the barycenters is approximately equal to the corresponding distance at the beginning of the convergence phase.
This precursor has the capability for a more accurate flare occurrence time prediction, which is based on the relationship between the durations of the divergence ($T_{D+F}$) and convergence phase ($T_{C}$) of the barycenters of the opposite polarity regions.
Next, K15 also found a linear relationship between the values of the maxima of the $WG_{M}$ ($WG^{max}_{M}$) and the highest flare intensity of an AR(s). They reported that if one can identify concurrently the two pre-flare behaviours discussed above, then flare(s) do occur in their sample. They have also shown that if one of the required pre-flare patterns is absent then a flare may not be expected \citep{Korsos2014,Korsos2016}.

Also, K15 investigated separately the single-flare case when only one energetic flare took place after $WG^{max}_{M}$ and cases when multiple flares erupted after reaching $WG^{max}_{M}$. In the 61 flare cases, the longest study period was 48 hrs from the moment of reaching $WG^{max}_{M}$ to the moment of first flare. The percentage difference ($WG^{\%}_{M}$) was calculated between the value of $WG^{max}_{M}$ and the first value of the $WG_{M}$ after the flare peak time ($WG^{flare}_{M}$). In brief, they found the following: if $WG^{\%}_{M}$ is over 54\%, no further flare of the same class or above would be expected; but, if $WG^{\%}_{M}$ is less than $\sim$42\%, further flare(s) of the same class is probable within about an 18-hour window. The longest time interval of a subsequent flare to occur was 18 hrs in the study samples. K15 suggested that these latter features may serve as practical additional flare alert tools.

In this study, we have generalised the application of the $WG_{M}$ method in two main ways. First, we have expanded the number of investigated ARs by taking into account not only ARs observed by the SOHO satellite but also those detected by the higher spatial and temporal resolution SDO (Solar Dynamics Observatory) mission. Second, we have extended the analysis to encompass GOES flare classes from as low as B-class to as high as X-class flares.
In Section \ref{Casestudy}, we briefly introduce and apply the $WG_{M}$ method to lower energetic flares, i.e. between B and M5 flares. In Section \ref{Analyses}, we present an extended statistical analysis of these higher number of AR cases and summarise our findings. In Section \ref{vizu}, we introduce a simple visualisation of our observational experience of the pre-flare behaviour of distance. Finally, we provide discussions of our results and draw conclusions in Section \ref{Conclusion}.

% case study

\section{Applying the $WG_M$  method} \label{Casestudy}

\subsection{Implementation of the $WG_M$ Method}

In 61 flare cases, K15 demonstrated that $WG_{M}$ could be successfully applied to help identify features preceding flares with classes above M5. 
Later, in \cite{Zheng2015} and \cite{Korsos2016}, the $WG_{M}$ method was applied to case studies of lower than M5 flare cases employing the SDO/HMI-Debrecen Data (also known as HMIDD, the continuation of the SDD) catalogue. In the present paper, we go further by enlarging the observational sample to include 6 ARs with B-class flares, 21 with C-class flares, 13 with M1-M5, and 30 additional ARs with flare events above M5 (see Appendix A).

The main reasons of the small number of the weaker than M5 flaring AR in the sample are as follows: (i) When the investigated strongest energetic flare class becomes lower and lower (e.g. it is below M5 or less) then there is an associated decreased chance of having this low-energy flare class to be the largest flare class of an AR. 
(ii) In principle, the HMIDD (2011-2014) database would be slightly more suitable to investigate the lower than M5-class flares because the temporal and spatial resolutions of this catalogue are better than that of the SDD. However, in case of these weaker flares, often, we simply cannot identify the two nearby opposite polarities of the AR in the HMIDD catalogue, so, the $WG_{M}$ method is not applicable.

By considering lower-energy flares, this expansion of the investigation of flare classes explores whether there could be a ``common'' physical mechanism underlying the flare process across all energy scales. However, it must be noted that our method does not give insight into which of the wide range of proposed flare models available in the literature is applicable. Our method points merely towards the idea that, regardless of model, there could be a ``common'' pre-flare features identified at photospheric level. 

The method is based on tracking changes of the solar surface magnetic configuration in ARs with the following five steps: 
\begin{enumerate}

\item During the entire investigated period it is required that the AR is located between $-70^{\circ}$ and $+70^{\circ}$ (occasionally, when the data permit $+75^{\circ}$, but not further) from the central meridian of the Sun. 
\item During the AR's disk passage, the largest intensity flare-class event of the AR is selected from the GOES flare catalog.
\item In order to acquire enough preceding data to identify the precursors, the occurrence of the associated strongest flare class could be no further than $\sim40^{\circ}$ east of the central meridian.
\item The $WG_{M}$ method is applied in a selected area of the AR. As an initial approach, the selected area is an entire $\delta$-spot of the AR where all umbrae are now taken into account for analysis. This assumption is based on the idea that the $\delta$-type sunspots themselves are observed and identified as the most probable places for the flare onset. A $\delta$-type spot contains opposite polarity umbrae surrounded by a common penumbra, therefore it has polarity inversion line(s) (PIL). It is also well known that solar flares often are related to PILs \citep[][]{Schrijver2007, Louis2015}. Furthermore, the umbrae are loci of high flux densities, so they are presumably the dominant components of the flare processes. However, it should be noted that the Debrecen sunspot catalogue does not always indicate the two close opposite magnetic polarities as a $\delta$-type spot. In this case, the selected area is a circle with a radius of 1.5$^{\circ}$$\pm 0.5$$^{\circ}$ around the barycentrum of the two closest opposite polarity umbrae. We introduced the circle in the studied samples of K15. The diameter of the circle is derived from the common amorphous shape penumbra of the opposite polarity umbrae approximated by a circle with a radius of 1.5$^{\circ}$$\pm 0.5$$^{\circ}$ in Carrington heliographic coordinates. Finally, the selected area is tracked and the evolution (e.g., emergence of new flux or flux cancellation) of umbrae are monitored. 
\item At the end, the $WG_M$ method is applied to the selected area. The evolution of the unsigned magnetic flux, the distance between the area-weighted barycenters of opposite polarities and the $WG_{M}$ are followed as outlined below: 

\begin{enumerate}[label=(\alph*)]
\item To establish that a behaviour is related to the upcoming flare rather than merely an insignificant fluctuation, (i) the relative gradient of the rising phase of $WG_{M}$ is 30\% and (ii) the relative gradient of the distance parameter of the converging motion is greater than 10\% for a period of at least 4 hrs. Furthermore, a maximum of 10\% deviation is allowed as the distance increases back to its original value that it had at the moment when the convergence phase started.
\item When the relevant pre-flare behaviour of the $WG_M$ proxy is identified as given in point a) above, then the $WG^{max}_{M}$ and $WG^{\%}_{M}$ can also be determined.  
\item Next, the relevant pre-flare behaviour of the distance parameter is marked with a parabolic curve. The parabolic curve is fitted from the starting time of convergence to the end of the divergence phase, where its minimum is the moment of reaching the closest position of the two barycenters. 

\end{enumerate}
\end{enumerate}

Let us comment on the errors and uncertainties of the $WG_{M}$ methods: The uncertainty in obtaining the distance parameter originates from the error of position measurements that is 0.1 heliographic degree, while measuring the area has 10\% error \citep{Gyori2011}. The mean error of the unsigned magnetic flux is 15\%. The magnetic field used here to calculate $WG_{M}$ is estimated from the values of umbral area and mean field strength recorded in the Debrecen (SDD and HMIDD) catalogues through the application of Eq. (1) in \cite{Korsos2014}. Therefore, the total calculated uncertainty of $WG_{M}$ is 20\%.

\subsection{Applying the $WG_M$ method to M-, C- and microflare cases  }

First, let us now demonstrate the technique of applying the $WG_M$ method to three representative examples where the flare classes are all {\it lower} than M5. The examples for analysis discussed here are: AR 11504 produced two low M-class flares; AR 11281 generated 3 C-class flares as the largest-class flare. Finally, AR 11967 is interesting because it hosted a known and identified microflare \citep{Yang2015}. In \cite{Yang2015}, observational evidence of X-shape magnetic reconnection before a microflare was introduced. The magnetic reconnection occurred at the topside edge of AR 11967 on February 3, 2014 07:15 UT. There, \cite{Yang2015} found that the X-shape reconnection process builds up of two types of reconnection: (i) First, two anti-parallel loops slowly reconnect, and, after the new loops were formed, they became stacked. This {\it slow reconnection} continued for several tens of minutes. (ii) The second type of reconnection, the {\it rapid reconnection}, took only about three minutes. During the {\it rapid reconnection}, the anti-parallel loops very quickly approached each other and reconnected. After the {\it rapid reconnection}, the former anti-parallel loops disappeared and new loops formed separately. 

\begin{figure} [h!]
\centering
\includegraphics[scale=1]{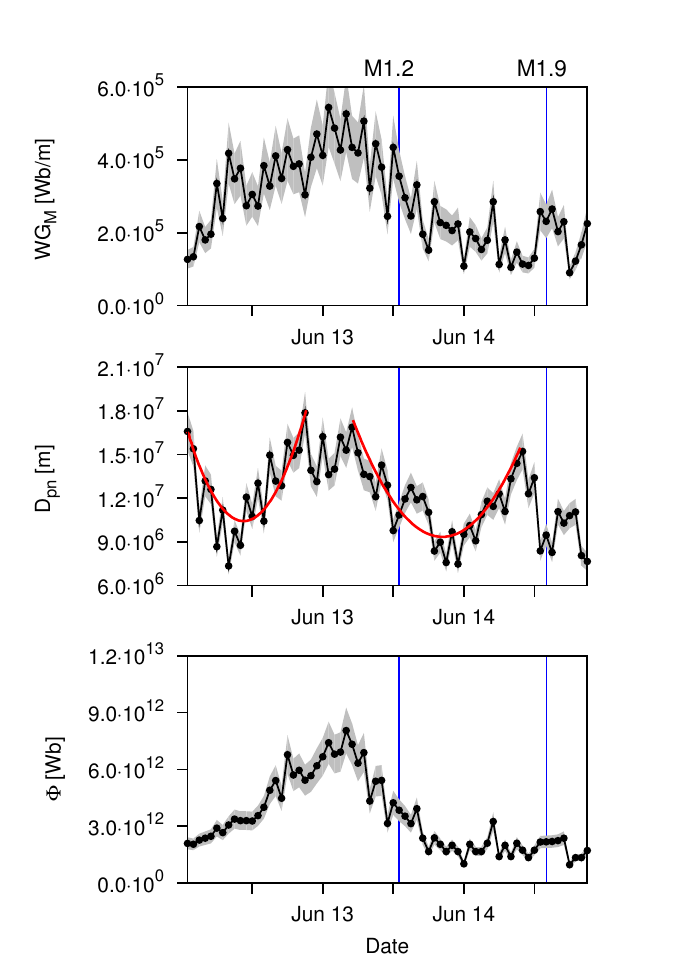} 
\put(-200,270){(a)}
\includegraphics[scale=0.32]{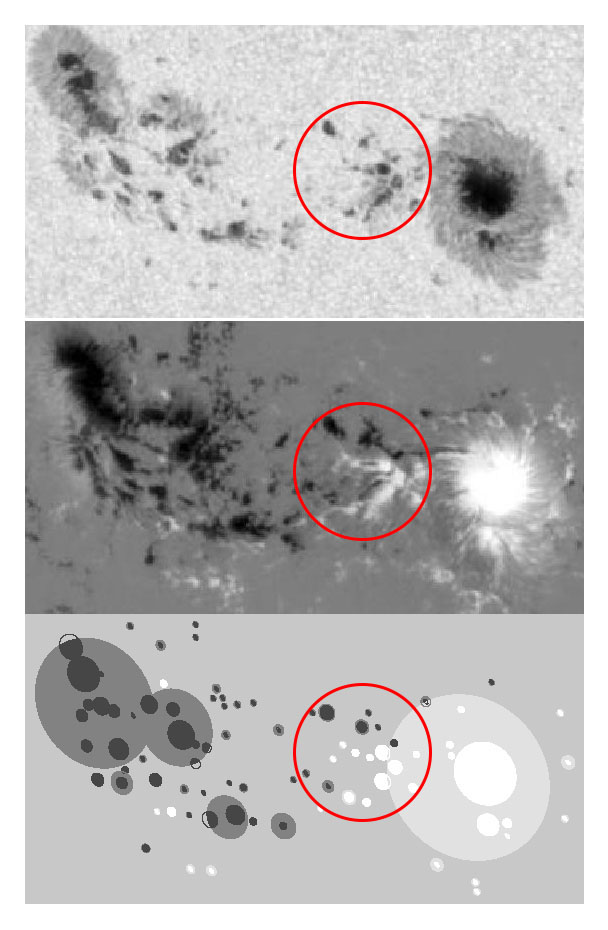} 
\put(-200,270){(b)}
\caption{ (a) Example for applying the $WG_{M}$  method to two smaller (M1.2 and M1.9) GOES M-class flares in AR 11504.  The top panel shows the $WG_{M}$, the middle panel plots the distance between the barycentres, and the bottom panel is a plot of the associated unsigned magnetic flux as a function of time. The M1.2 and M1.9 flares are indicated by (blue) vertical lines where the value of the $WG^{flare}_{M}$ is taken. Note the U-shapes (red parabolae) in the middle panel that are key flare precursors of the $WG_{M}$ method. The error is marked with shaded grey. (b) Top/middle and bottom panels are the related intensity/magnetogram maps and the associated Debrecen data catalogue representation of  synthetic polarity drawing of AR at 01:59 on 13 June 2012. The red circle shows which umbral area and mean field strength are taken to calculate $WG_{M}$.}
\label{AR11504}
\end{figure}

\begin{figure} [h!]
\centering
\includegraphics[scale=1]{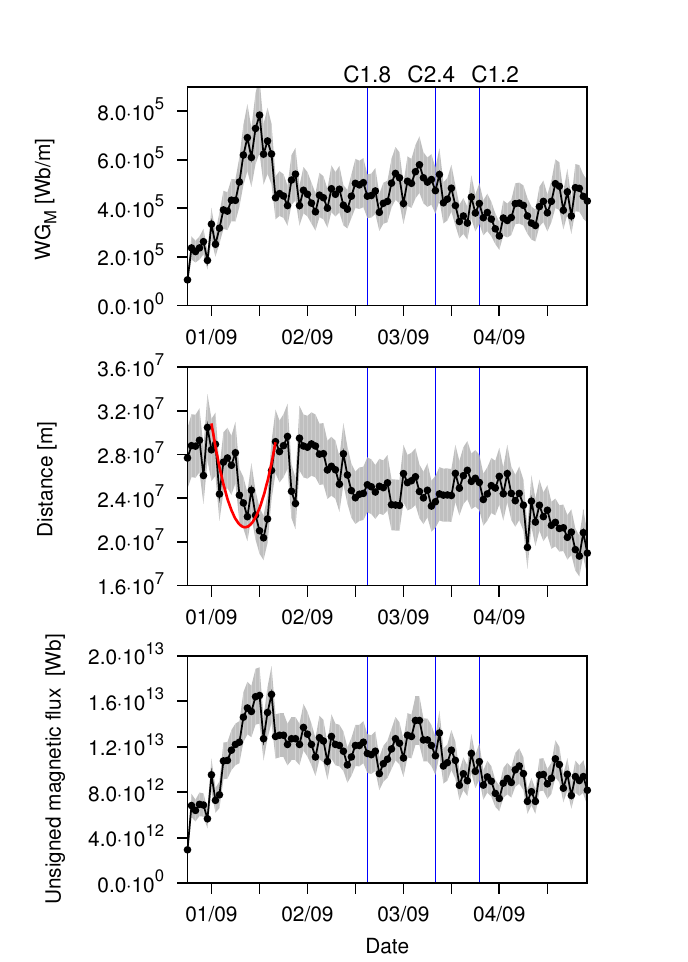} 
\put(-200,270){(a)}
\includegraphics[scale=0.28]{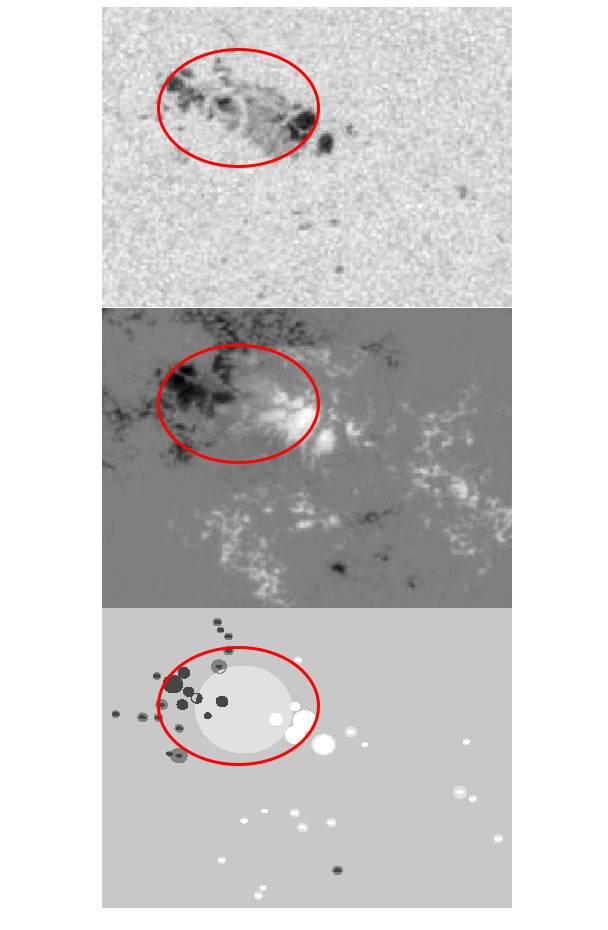} 
\put(-180,270){(b)}
\caption{(a) Representative example for applying the $WG_{M}$ method to GOES C-class flares in AR 11281. (b) Top/middle and bottom panels are the associated intensity/magnetogram maps and the Debrecen catalogue representation of the associated synthetic polarity drawing at 00:59 on 03 September 2011.  }
\label{AR11281}
\end{figure}

\begin{figure} [h!]
\centering
\includegraphics[scale=1]{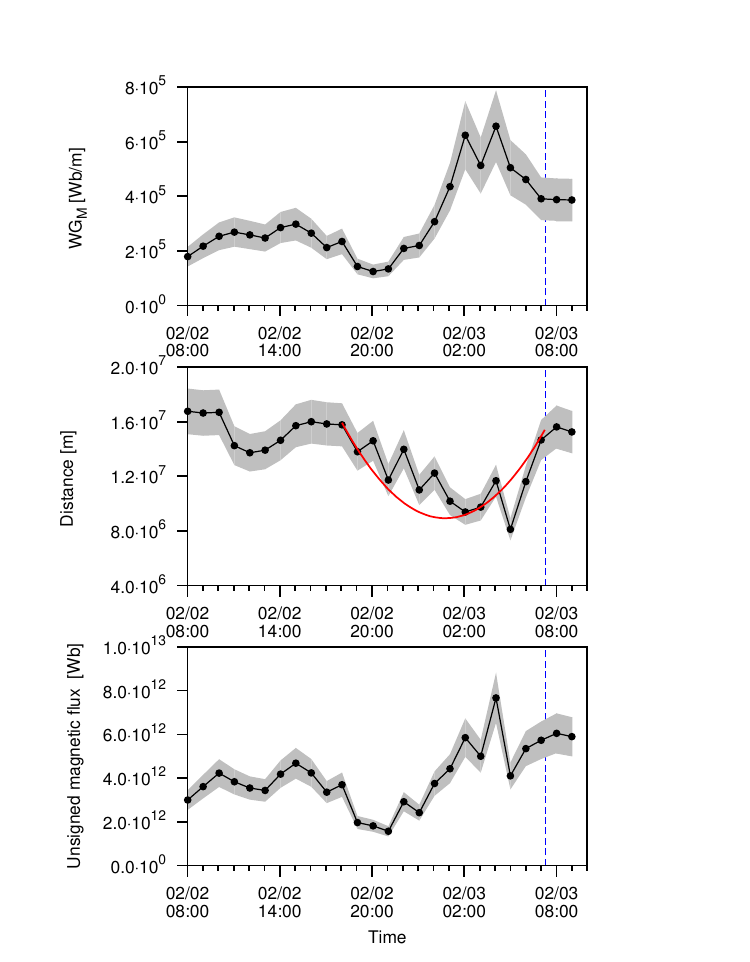} 
\caption{ Evolution of the pre-flare indicators of the $WG_M$ method, similar to those of Fig. 2 but for a B-class flaring event in AR 11967. The estimated error is marked by the shaded grey envelope. }
\label{AR11967}
\end{figure}

The resulting diagrams of the $WG_{M}$ analysis of ARs 11504, 11281, and 11967 are shown in Figs. \ref{AR11504}a, \ref{AR11281}a and \ref{AR11967}, respectively. In Figs. \ref{AR11504}a--\ref{AR11967}, we depict the pre- and post-flare evolution of $WG_{M}$ (top panel) and we also plot the distance between the barycentres of opposite polarities as a function of time (middle panel). In the middle panels of Figs. \ref{AR11504}a--\ref{AR11967}, the relevant pre-flare behaviour of the distance parameter is marked with red parabolic curves. In the bottom panels of Figs. \ref{AR11504}a--\ref{AR11967}, we show the temporal variation of the unsigned magnetic flux in selected area. Figures~\ref{AR11504}b--~\ref{AR11281}b visualise the investigated area by red ellipses in their white-light appearance (upper panel), the corresponding magnetogram (middle panel) and the corresponding synthetic polarity drawing from the Debrecen sunspot data catalogue (bottom panel) from which umbral area and mean field strength are taken to calculate $WG_{M}$.

Let us first investigate the case where the largest intensity flare class was a weak M-class. AR 11504 produced an M1.2 flare on June 13, 2012 13:17 UT and a further M1.9 flare on June 14, 2012 14:35 UT. 
In Figure \ref{AR11504}a, we see that the $WG_M$ increases to a maximum value ($WG^{max}_{M}$=$0.55$$\cdot$$10^6$ Wb/m), followed by a less steep decrease which ends with an M1.2 ($WG^{flare}_{M}$=$0.35$$\cdot$$10^6$ Wb/m) flare and is succeeded by another M1.9 ($WG^{flare}_{M}$=$0.23$$\cdot$$10^6$ Wb/m ) energetic flare (top panel). 
The conditions specified by step 5a are satisfied by the pre-flare behaviour of the $WG_{M}$, therefore we can dismiss the idea that this behaviour is unconnected to the flare. Moreover, the percentage difference ($WG^{\%}_{M}$) is only 34\% in the first flare case, which is less than 42\% (as defined by the criterion given by K15), therefore we expected more flares to follow, which did indeed happen. After the second flare, the $WG^{\%}_{M}$ is 61\%, so further flares were not expected. 
Regarding the distance parameter, the two convergence-divergence phases are evident. The first one is before the M1.2 flare where the $T_{C}$ is 9 hrs and $T_{D+F}$ is 34 hrs. The second one occurred before the M1.9 flare where $T_{C}$ is 14 hrs and $T_{D+F}$ is 19 hrs. It is worth mentioning that the two convergence phases of the barycenters had a duration longer than 4 hrs; the relative gradient of the first decreasing phase is 56\% and for the second one it is 54\%.

Let us now introduce a representative example of the analysis of a C-class flare of this by investigating, AR 11281. This AR was the cradle to the following 3 C-class flares: C1.8 on September 2, 2011 15:16 UT, C2.4 and C1.2 on September 3, 2011 07:56 and 20:10 UT, respectively.  
In Fig. \ref{AR11281}a, we recognise the following pre-flare properties of the $WG_M$ and the distance: (i) The rising phase and a maximum value of the $WG_M$ ($WG^{max}_{M}$=$0.78$$\cdot$$10^6$ Wb/m) is followed by a less steep decrease which ends with C1.8 ($WG^{flare}_{M}$=$0.45$$\cdot$$10^6$ Wb/m), C2.4 ($WG^{flare}_{M}$=$0.51$$\cdot$$10^6$ Wb/m) and after that with the C1.2 ($WG^{flare}_{M}$=$0.36$$\cdot$$10^6$ Wb/m) energetic flares. 
The conditions of point 5a are satisfied, therefore the pre-flare behaviour can be confidently attributed to the flare. (ii) The convergence-divergence feature of the barycentric distance prior to the first C-class flare are also evident. The duration of the convergence phase of the distance is 13 hrs and the gradient is 26\%. The first flare occurred 30 hrs later, measured from the moment of the closest position of the two opposite polarity barycenters. The second C-class flare occurred approximately 17 hrs after the first C-class flare. The final C1.2 flare occurred 12 hrs after the second C2.4 flare.
Let us now briefly investigate the percentage differences of the three flares. The $WG^{\%}_{M}$ is 42\% after the first C-class flare (C1.8). The $WG^{\%}_{M}$ is 35\% after the C2.4 and 54\% after the last C-class occurrence from the previous $WG^{max}_{M}$. We conclude that one should indeed expect flare(s) after the first C1.8 flare, and that one should not expect further same class flare(s) after the last C-class, which is what happened. Because there was only one clear U-shape flare precursor, we could not say anything about how many same-class flares will follow the first C flare.

In Figure \ref{AR11967}, for AR 11967 we analysed  the same area in the HMIDD catalogue as was analysed from HMI line-of-sight magnetograms in Fig. 1 of \cite{Yang2015}.

One can indeed recognise the increasing and decreasing phase of the $WG_{M}$ before the microflare. The maximum value of the $WG_M$ is $0.65$$\cdot$$10^6$ Wb/m and the value of the $WG^{flare}_{M}$ is $0.38$$\cdot$$10^6$ Wb/m. The $WG^{\%}_{M}$ is 42\% after the maximum of the $WG_M$. Unfortunately, we cannot say whether a further flare occurred because we do not have any later observations from this area. Next, the convergence and divergence phases of the distance are also identifiable: we emphasise this with a red parabola in the middle panel of Fig. \ref{AR11967}. Here, the duration of the observed $T_{C}$ is 7 hrs, with 40\% decreasing of the distance and $T_{D+F}$ is 6 hrs. Based on the required conditions and steps outlined in 1-5 above, these two pre-flare behaviours can be classified as true precursors of the microflare. 
Note that there may be another typical pre-flare behaviour of the $WG_{M}$ and distance between 02/02 08:00 and 02/02 18:00. Although, based on 5(a), the pre-flare behaviour of the $WG_{M}$ could be a precursor, but, the pre-flare behaviour of distance does not qualify as a precursor because the decreasing time is only one hour.
In summary, as Fig. \ref{AR11967} demonstrates, it is clear that even microflares seem to show the precursors of flaring identified by the horizontal magnetic gradient method. 

From these three sample studies presented (Figs. \ref{AR11504}--\ref{AR11967}) and, supported by analyses of the entire ensemble data, we propose that the pre-flare behaviour of $WG_{M}$ and the distance of the area-weighted barycentre of opposite polarities may be present widely and may be indispensable before the associated reconnection and/or flaring process. If the conjecture of pre-flare behaviour is proven to even more solar data than the current ensemble of 127 AR cases, this will certainly give us a greater statistical significance for understanding the underlying physics.

%statistic

\section{Statistical analyses of $WG_M$ method on the extended data} \label{Analyses}

\begin{figure} [h!]
\centering
	\includegraphics[scale=0.7]{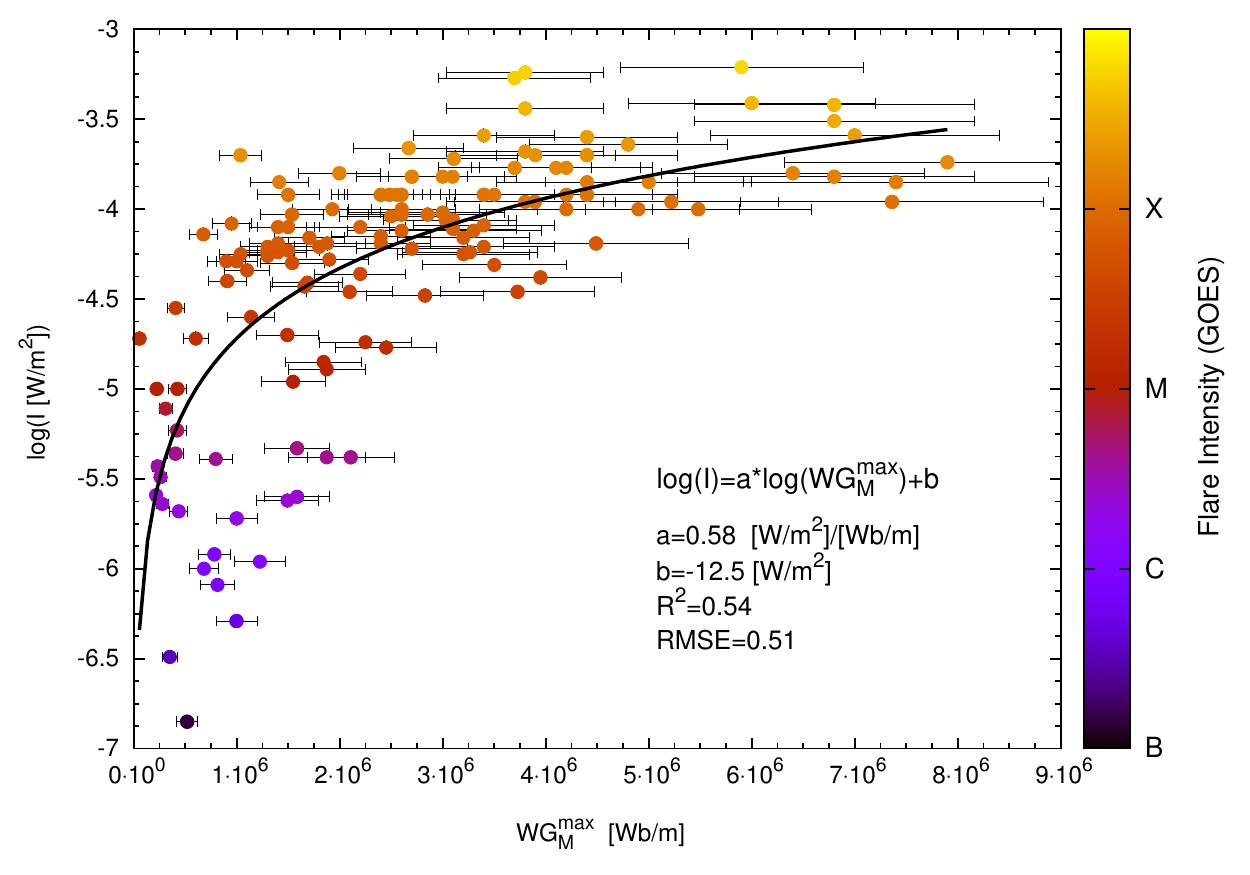} 
\caption{The log value of GOES flare intensity ($log(I)$) as function of the maximum $WG_{M}$ ($WG^{max}_{M}$). The estimated errors are also given in the lower right corner.}
\label{Maximum}
\end{figure}

Our aim is to analyse the photospheric precursors of flares of a 127 strong set of AR from SDD and HMIDD.
First, let us focus on the relationship between the log value of largest intensity flare of an AR ($log(I) $) and the preceding maximum of the $WG_{M}$ (see Fig. \ref{Maximum}).
We have found a logarithmic dependence between the $log(I)$ and the $WG^{max}_{M}$. The correlation coefficient of the fitted logarithmic function is $R^{2}=0.54$ and is an indicator for a reasonable functional fit to the data. The root mean square error (RMSE) of $log(I [W/m^{2}]$) is 0.51.

Next, the $WG_{M}$ method reveals further important connections between the proposed precursors and the associated flare properties. K15 showed that, for large flares, there is a relationship between the duration of the converging motion ($T_{C}$) and the sum of the duration of the diverging motion of the barycenters of opposite polarities together with the remaining time until flare peak ($T_{D+F}$). 
The question is then whether this relationship is also valid in the extended data studied here. In other words, it is of interest to establish whether this relationship found for flares above M5 remains for less energetic flares, i.e. below M5 down to C-class or microflares. 

\begin{figure} 
\centering
	\includegraphics[scale=0.63]{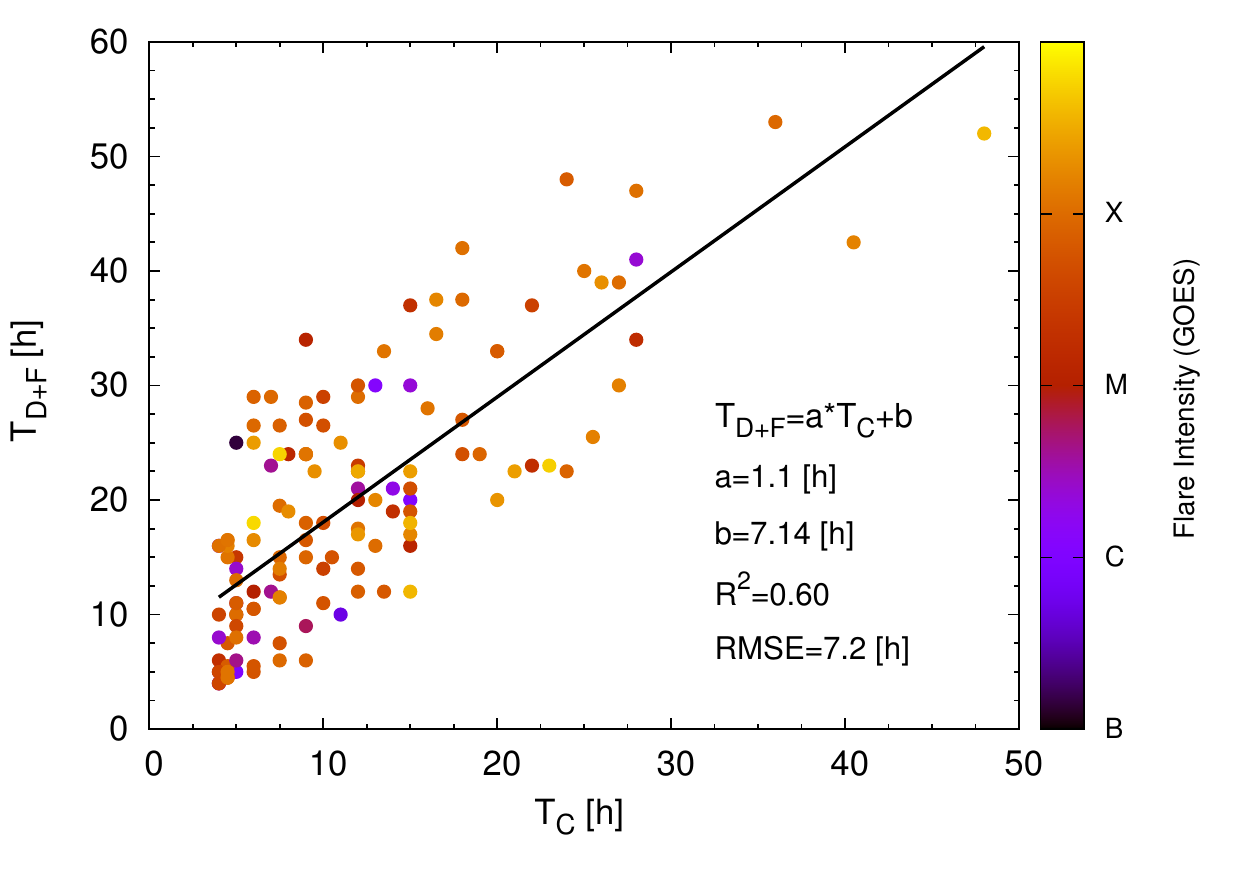} 
	\includegraphics[scale=0.63]{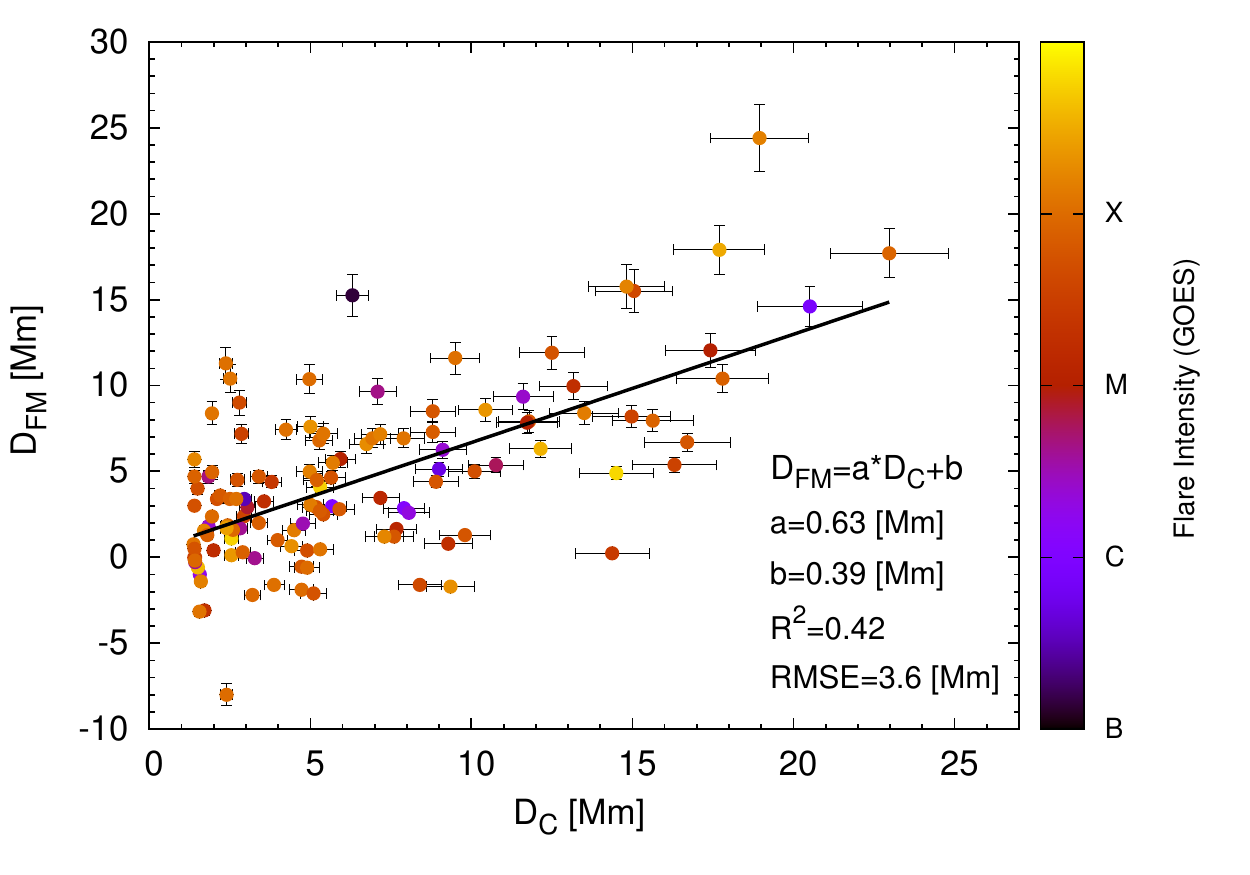} 
\caption{  {\it Left}: Relationship between the durations of converging motion and the duration from the moment of time of closest position up to the first flare occurrences. {\it Right}: Relationship between difference of the distances between the barycenters at the start of the convergence phases and at the closest approach ($D_C$) and the distance between the point of closest approach to the position of first flare occurrence ($D_{FM}$) at photospheric level. The estimated errors are given in the lower right corner.}
\label{pushpull}
\end{figure}

Figure \ref{pushpull} (left panel) gives a further insight into the relation between these physical quantities by plotting the elapsed time between the start of the divergence phase and the flare peak as a function of the duration of converging motion. The linear relationship found may possess the capability to estimate an approximate occurrence time of the associated flare. $R^{2}$ of the fitted linear function is 0.60 indicating a moderate correlation. By identifying the start of the divergence phase of the barycenters of opposite polarity, one may predict the time of first flare occurrance with an estimated error of 7.2 hrs.
We also investigated whether there is a correlation between the duration of converging-diverging motion and the flare intensity, but we were unable to conclude any statistically significant relationship.

Figure \ref{pushpull} (right panel) shows the linear correlation between the distance from the starting point of the converging phase to the point of the closest approach ($D_C$) and the distance between the point of closest approach to the position of the first flare occurrences ($D_{FM}$). 
 The linear fit between $D_C$ and $D_{FM}$ may provide another practical tool for estimating the spatial location of the flare. Here, the $R^{2}$ of the linear regression is only 0.42 which means that the correlation is moderate. The RMSE is 3.6 Mm. Again, we cannot report any statistically significant relationship between the distance values and flare intensity.
However, it is worth mentioning that the expected occurrence time and estimated location could both reinforce the search for a more reliable flare prediction.

Last, but not least, we carried out an analysis similar to that of K15 to estimate the corresponding probability thresholds and have found reassuring results confirming the earlier findings. Namely, if the $WG^{\%}_{M}$ is over 55\%, no further energetic flares are expected; but, if the $WG^{\%}_{M}$ is less than $\sim$40\%, a further flare is probable within approximately 18 hrs. If the $WG^{\%}_{M}$ is between 40\% and 55\%, one cannot make a reliable prediction of whether additional flares will/will not take place. In summary, therefore, these properties of the $WG_{M}$ method may serve as practical flare watch alert tools across a wide range of the flare energies, subject of course to the conditions outlined in Section \ref{Casestudy}.

%vizulition

\section{Visualisation of pre-flare behaviour of the distance parameter} \label{vizu}

\begin{figure}
\centering
\subfigure{\includegraphics[width=0.35\columnwidth]{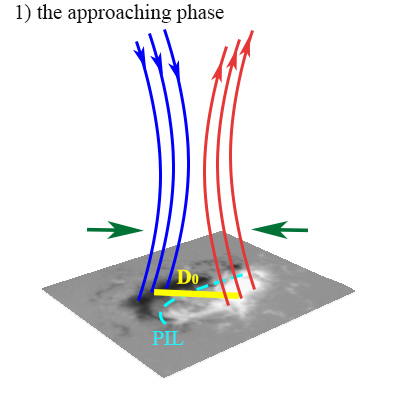}}
\subfigure{\includegraphics[width=0.35\columnwidth]{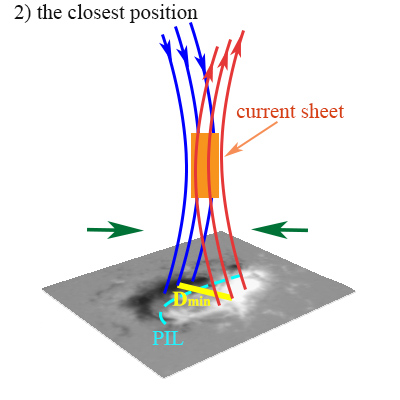}}
\subfigure{\includegraphics[width=0.35\columnwidth]{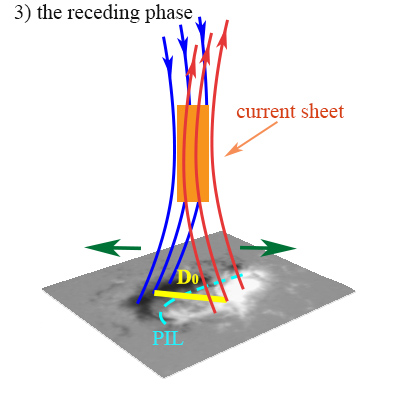}}
\subfigure{\includegraphics[width=0.35\columnwidth]{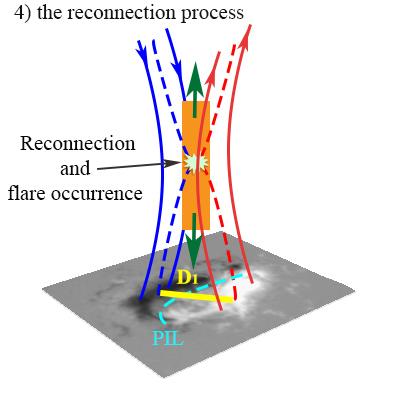}}

\caption{\label{reconnection} Figure demonstrating the process (1) when two opposite polarities of initial barycenter separation distance $D_{0}$ are convergence; (2) the two opposite polarities are at their closest and a current sheet starts forming, where $D_{Min}$ is the minimum distance, with $D_{0}$ $>$ $D_{Min}$; (3) the two opposite polarities are divergence from their closest distance, $D_{Min}$, back to $D_{0}$, and the associated current sheet is still developing above the photosphere; (4) reconnection takes place and a flare occurs above the polarity inversion line (PIL). After the distance between the polarity barycentres returns to the initial separation, $D_{0}$, during a further evolution, this distance can now either decrease or increase (i.e. $D_{1}$ $<$  $D_{0}$ or $D_{1}$ $>$  $D_{0}$). $D_C$=$D_{0}$-$D_{Min}$ in Fig. \ref{pushpull} (right panel).
}

\end{figure}

In this section, inspired by laboratory experiments, we introduce a simple visual interpretation of the observed pre-flare behaviour of the area-weighted barycenters of opposite polarities prior to the reconnection process.
The process of magnetic reconnection in the solar atmosphere is mostly studied either using space-based observations or theoretical (e.g. numerical or analytical) modelling. However, laboratory experiments may also yield some interesting insight and impetus. A good example is the series of experimental studies by e.g. \cite{ Yamada1999} and \cite{Yamada2010}. \cite{Yamada1999} investigating the physics of magnetic reconnection in a controlled laboratory environment. In these experiments, reconnection is driven by torus-shaped flux cores which contain toroidal and poloidal coil windings. 
Two types of reconnection modes were found, according to whether the poloidal field coil current increased or decreased. When the poloidal coil current increased then the poloidal fluxes increased as well and plasma was pushed toward the X-point. This reconnection process is called the {\it push mode}. On the other hand, when the poloidal current decreased, the associated decreasing poloidal flux in the common plasma was pulled back toward the X-point, a reconnection process known as the {\it pull mode}. They found that the {\it push mode} occurs more rapidly than the {\it pull mode}. 

Besides the extensive modelling in the literature, the experiments by \cite{Yamada2010} have been a direct drive to solar observational studies of the process of magnetic reconnection. For example, in K15, it was found that the area-weighted barycentres of two opposite magnetic polarities of an AR in the investigated area first approach each other, reach their minimum distance followed by a divergence phase. Most surprisingly, the flare occurrence(s) take place when the growth of the distance becomes large enough and it approaches the value it had at the beginning of the convergence phase (for the actual details see their middle panel of Figs. \ref{AR11504}a--\ref{AR11967}). K15 have shown, using 61 samples from the SOHO (Solar and Heliospheric Observatory) era, that there is never a large flare occurrence when the barycenters are closest. Occasionally, though, smaller so-called precursor flares may take place. The divergence phase was found to continue until the distance increased back to about its original value, i.e. to the level of separation when the convergence phase started. The most energetic flares were found to happen after the divergence phase, and for the flare occurrence time a statistical relationship was established in terms of the duration of the convergence/divergence phases (see left-hand side of Fig. \ref{pushpull}).

In Figure \ref{reconnection}, we introduce a simple visualisation of the pre-flare behaviour. First, the two opposite magnetic polarities start convergence (panel 1), with an initial barycentric distance of $D_0$. When the barycentres reach their closest position (i.e. the separation is $D_{min}$), a thin current sheet begins to form between the opposite polarity field lines (or sheets) but there is no reconnection yet (see panel 2). After the minimum distance stage, the two opposite polarities begin to recede from each other and the separation at photospheric level increases back to the about the same level of separation when the converging process started (see panel 3), with barycentric separation distance of $\sim$$D_0$. The current sheet is still forming during the divergence phase above the photosphere. Finally, reconnection takes place, however, well {\it after} the moment when the photospheric distance between the area-weighted centres of polarity is at its about the same value of what it had at the beginning of the convergence phase (see panel 4), with a barycentric distance $D_1$. During the process of magnetic reconnection, the magnetic field lines rearrange according to the yet unknown key principles of reconnection in the highly stratified lower solar atmosphere. This rearrangement is accompanied with a sudden energy release, e.g. flare eruption, where the energy of eruption was stored in the stressed magnetic fields. 

A further possible explanation of our empirical finding may be that the actual convergence phase is caused by bipolar flux emergence between the two barycenters at the area of the PIL that eventually brings the barycenters closer. Next, the divergence may be caused by the strong shearing motion between the opposite polarities \citep{Ye2018}.

%conclusion

\section {Conclusions} \label{Conclusion}

Most flare forecasting models attempt to predict flare probability \citep[see, e.g.,][and references therein]{Georgoulis2012,Georgoulis2013, Barnes2016}. Many of these flare forecast studies focus on a predictive time window of 6, 12, 24 and 48 hrs \citep[see, e.g.,][and references therein]{Al-Ghraibah2015, Benz2017}. In K15, the concept of {\it the weighted horizontal gradient of the magnetic field}, $WG_{M}$, was introduced where all umbrae were taken into account in the selected $\delta$-spot for analysis. Initially, the $WG_{M}$ method was tested on data available from the SOHO era only. 

Here, we carry out an extended statistical analysis of these photospheric precursors of pre-flare dynamics on a larger sample of $\delta$-spots observed, including not only those contained by SDD but also by those found in HMIDD. The main motivation is to further develop, improve and confirm the applicability of two parameters (the $WG_{M}$ and barycentric distance parameters) introduced. Inspired by the results of \cite{Zheng2015} and \cite{Korsos2016}, we expanded the statistical sample of flares to be investigated below the M5-class down to B-class microflares, therefore offering an over-arching view of the applicability of the $WG_{M}$ method for a wider energy spectrum of flares. An answer is searched for to the question, do smaller flares display the same predictive pre-flare features as their stronger cousins. In the present work, we have outlined the case for the affirmative answer.

In Section \ref{Casestudy}, we introduced three representative $\delta$-type ARs, AR 11504, AR 11281 and  AR 11967 for presenting characteristic sample studies from our extended dataset. These are typical ARs for less significant GOES energetic flare classes. Based on Debrecen Heliophysical Observatory catalogues, in all of the observed $\delta$-spot of the 127 ARs, we have identified the two pre-flare patterns established earlier in K15: (i) {\it the pre-flare behaviour pattern of $WG_{M}$}: a rising phase, a maximum and a gradual decrease prior to flaring and (ii) the  pre-flare evolutionary pattern of the distance of the area-weighted barycentres during the converging and diverging motion from the minimum distance value of the area-weighted barycenters of opposite polarities until the flare occurrence. Furthermore, we have set out empirical conditions that the $WG_{M}$ and the barycenter distance parameters have to satisfy to qualify as being precursive of a flare rather than an unrelated fluctuation.

After identifying the pre-conditions, we have also investigated the relationship between the intensity of flares from $\delta$-spots in terms of the $WG^{max}_{M}$ (see Fig \ref{Maximum}). We have always focussed on the largest intensity flare ({\it I}) which has occurred in the given AR after reaching $WG^{max}_{M}$. By extending the flare samples down to B-class, we found a logarithmic relationship between the $log(I)$ of the investigated ARs and $WG^{max}_{M}$. This relationship may provide a tool to estimate the $log(I)$ of the expected flare with $\pm$ 0.51 uncertainty from the measured $WG^{max}_{M}$. 
 
In K15, we found a linear relationship between the duration of the converging motion and the time elapsed from the moment of minimum distance until the flare peak. Our extended statistical sample from SDD and HMIDD data from B-class to X-class flares, again, confirms this linear relationship. Therefore, we propose that if one can reliably identify the moment 
when the barycenter distance in a $\delta$-spot begins to grow again then one is able to estimate the occurrance time of the flare with $\pm$ 7.2 hrs of uncertainty. Furthermore, we also investigated the connection between the length of distance from the starting point of the converging phase to the point of closest approach ($D_C$) and the distance between the point of closest approach to the position of the first flare ($D_{FM}$) between the area-weighted barycentres. The linear relationship, with an estimated error of 3.6 Mm, found between $D_C$ and $D_{FM}$ may help to identify the region where the flare occurrence may be expected. 
So, the expected time of flare occurrence and its predicted location could serve as combined tools for flare warning.

We have also searched for phenomenological clues for differences for predicting the number of flares expected from $\delta$-spots after the horizontal magnetic gradient reaches $WG^{max}_{M}$, during its decreasing phase. In K15, for flares stronger than M5, the conclusion was: if the percentage difference ($WG^{\%}_{M}$) between the value of the $WG^{max}_{M}$ and the first value of the $WG_{M}$ after the flare peak ($WG^{flare}_{M}$) is over 54\%, no further energetic flare(s) may be expected; but, if the percentage difference is less than about $\sim$42\%, further flaring is likely within the following 18 hrs. In the present study, we have revisited the estimated probability of further flares during the descending phase of the $WG_{M}$ after its maximum. We found encouraging results extending the initial findings of K15 to a wider flare energy range, namely: if the percentage difference ($WG^{\%}_{M}$) is over 55\%, no further energetic flare(s) may be expected; but, if $WG^{\%}_{M}$ is less than $\sim$40\%, further flaring is probable within about 18 hrs. The importance of this empirical result is that it could be a further auxiliary tool for indicating the properties of imminent flares from $\delta$-spots.

\section{ Acknowledgements} 
     MBK is grateful to the University of Sheffield and the Hungarian Academy of Sciences for the supports received. MBK also acknowledges the open research program of CAS Key Laboratory of Solar Activity, National Astronomical Observatories, No. KLSA201610. MBK and RE acknowledge the CAS Key Laboratory of Solar Activity, National Astronomical Observatories  Commission for Collaborating Research Program for support received to carry out part of this work. RE acknowledges the CAS Presidents International Fellowship Initiative, Grant No. 2016VMA045.  RE is also grateful to Science and Technology Facilities Council (STFC, grant number ST/M000826/1) UK and the Royal Society for enabling this research.  SY is supported by the National Natural Science Foundations of China (11673035, 11790304). All authors thank Christopher Nelson and Matthew Allcock for fruitful discussions. Last but not least, the authors also would like to thank the two anonymous referees and the Guest Editor for their helpful comments received during the peer-review evaluation process. 
 
\bibliography{swsc.bib}

\appendix
\section{List of investigated ARs}

The first column is the NOAA AR number. The second column is the largest flare-class during the AR's disk passage (M5$<$ denotes classes between M5-M9.9 and M1$<$ stands for M1-M4.9). The third and fourth columns include the starting and finishing moments and the corresponding locations of the AR analysis.

\begin{table} [ht!]
\centering
\begin{tabular}{|c|c|cc|cc|}

\multicolumn{2}{c|}{}          &              \multicolumn{4}{c|}{\cellcolor{Gray}{1997}} \\
\hline 
8088&	M5$<$&	22/09 00:00	&	S28E53	&	24/09 23:59	&	S28E10	\\
\hline 
 8100&	X&	03/11 00:00	&	S19W12	&	04/11 23:59 	&	S21W39	\\
\hline 
\end{tabular}
\end{table}

\begin{table} [ht!]
\centering
\begin{tabular}{|c|c|cc|cc|}

\multicolumn{2}{c|}{}          &              \multicolumn{4}{c|}{\cellcolor{Gray}{1998}} \\
\hline   
8210&\textcolor{white}{1}X\textcolor{white}{$<$}&		01/05 00:00	&	S17W03	&	03/05 23:00  &	S17W36	\\
\hline 
\end{tabular}
\end{table}

\begin{table} [ht!]
\centering
\begin{tabular}{|c|c|cc|cc|}
\cline{3-6}
\multicolumn{2}{c|}{}          &              \multicolumn{4}{c|}{\cellcolor{Gray}{1999}} \\
\hline   
8485&	M5$<$&	 	14/03 00:00	&	N23E00	&	16/03 23:59	&	S14W43	\\
\hline 
8647	& X	&			01/08 17:00	&	S18W18	&	04/08 14:00	&	S18W63	\\
\hline   
8771&	X	&23/11 16:00	&	S15W20	&	27/11 13:00	&	S14W71	\\
\hline   
8806	&	M5$<$	&	20/12 00:00	&	N24E48	&	25/12 00:00	&	N24W18	\\
\hline 
\end{tabular}
\end{table}

\begin{table} [ht!]
\centering
\begin{tabular}{|c|c|cc|cc|}
\cline{3-6}
\multicolumn{2}{c|}{}          &              \multicolumn{4}{c|}{\cellcolor{Gray}{2000}} \\
\hline   
8882&	X&	  01/03  00:00	&	S18W31	&	02/03 23:59	& S16W60	\\
\hline   
8910&	X&	 	 19/03  00:00	&    N11W10	&	22/06 23:59	&	N13W61	\\
\hline   
9026&	X&	 	 04/06  00:00	&	N20E48	&	08/06 23:59	&	N22W17	\\
\hline   
9077&	X&	 	 10/07  00:00	&	N18E55	&	14/07 23:59	&	N18W09	\\
\hline   
9090&	M5$<$&	  20/07  00:00	&	N11E32	&	 21/07 10:00	&	N12E05	\\
\hline   
9087&	M5$<$&	 	  18/07  00:00	&	S12E28	&	 19/07 23:59	&	S12E13	\\
\hline   
9097&	M5$<$&	  23/07  00:00	&	N06E25	&	 25/07 23:59	&	N08W15	\\
\hline   
9165&	M5$<$&	 	 15/09  00:00	&	N13E14	&	19/09 10:00	&	N14W40	\\

\hline   
\end{tabular}
\end{table}

\begin{table} [ht!]
\centering
\begin{tabular}{|c|c|cc|cc|}
\cline{3-6}
\multicolumn{2}{c|}{}          &              \multicolumn{4}{c|}{\cellcolor{Gray}{2001}} \\
\hline   
9368&	M5$<$&	 07/03   12:00	&	N25W15	&	08/03  23:59	&N26W33	\\
\hline   
9393&	X&		 26/03   00:00	&	N20E39	&	02/04  23:59	&N16W70	\\
\hline   
9415&	X&	 	 05/04   00:00	&	S21E60	&	14/04  23:59	&S22W72	\\
\hline   
9433&	M5$<$&	 	23/04     00:00	&	N17E26	&	29/04     23:59	&N17W50\\
\hline   
9503&	M5$<$&	 	21/06    00:00	&	N16W20	&	22/06    23:59	&N17W46\\
\hline   
9511&	X&	 	 22/06   14:00	&	N10E30	&	23/06  23:59	&N10E00\\
\hline   
9601&	M5$<$&	 	04/09    00:00	&	N14W06	&	05/09    23:59	&N14W38\\
\hline   
9608&	M5$<$&	 	14/09    00:00	&	S25W33	&	17/09    23:59	&S28W75\\
\hline   
9628&	M5$<$&	 	23/09     00:00	&	S17E25	&	27/09    23:59	&S18W01\\
\hline   
9632&	X&		22/09     00:00	&	S17E56	&	24/09    23:59	&S19E06\\
\hline
9661&	X&	 	13/10   00:00	&	N14E55	&	19/10   23:59	&N16W35\\
\hline   
9672&	X&	 	23/10   00:00	&	S18E13	&	25/10   23:59	&S18W27\\
\hline   
9684&	X&	 	01/11   00:00	&	N06E29	&	04/11   23:59	&N05W28\\
\hline   
9704&	X&	17/11   00:00	&	S18E41	&	22/10   23:59	&S18W38\\
\hline   
9727&	M5$<$&	 	09/12   00:00	&	S22E03	&	12/12   23:59	&S21W52\\      
\hline   
9733&	X&	 	10/12   00:00	&	N14E58	&	18/12   23:59	&N13W65\\   
\hline   
9742&	M5$<$&22/12   00:00	&	N10W03	&	26/12   23:59	&N12W68\\      
\hline 
\end{tabular}
\end{table}

\begin{table} [ht!]
\centering
\begin{tabular}{|c|c|cc|cc|}
\cline{3-6}
\multicolumn{2}{c|}{}          &              \multicolumn{4}{c|}{\cellcolor{Gray}{2002}} \\
\hline   
9773&	M5$<$&	 	 08/01   14:00	&	N12E17	&	09/01  23:59	&N14W05\\
\hline   
9866&	M5$<$&	 	 12/03   00:00	&	S10E43	&	14/03  23:59	&S10E07	\\
\hline   
10017&	X&		 02/07   00:00	&	S19W37	&	03/07  23:59	&S18W63	\\
\hline   
10044&	M5$<$&	 	 25/07   17:00	&	S20E34	&	26/07  23:59	&S21E17	\\
\hline  
10069&	X&	 	 14/08   00:00	&	S07E50	&	21/08  06:00	&S08W50	\\
\hline       
10226&	M5$<$&	 	 16/12   00:00	&	S28E25	&	20/12  23:59	&S28W41	\\
\hline                   
\end{tabular}
\end{table}

\begin{table} [ht!]
\centering
\begin{tabular}{|c|c|cc|cc|}
\cline{3-6}
\multicolumn{2}{c|}{}          &              \multicolumn{4}{c|}{\cellcolor{Gray}{2003}} \\
\hline   
10314&	X&		 15/03   11:00	&	S14W05	&	18/03  23:59	&S16W52\\
\hline           
10338&	M5$<$&	 	 22/04   11:00	&	N18W10	&	26/04  22:00	&N18W70\\
\hline                  
10365&	X&	 	 25/05   00:00	&	S08E11	&	30/05  23:59	&S07W59\\
\hline                  
10375&	X&		 06/06   00:00	&	N12E24	&	11/06  23:59	&N12W62\\
\hline                  
10484&	X&	  20/10   00:00	&	N06E53	&	28/10  23:59	&N03W68\\
\hline                   
 10486&	X&	 25/10   00:00	&	N06E53	&	02/11  23:59	&N03W68\\
 \hline                   
 10488&	X&	  28/10   00:00	&	N09E09	&	03/11  12:00	&N08W74\\
\hline                   
 10501&	M5$<$&	 	 15/11   00:00	&	N04E61	&	21/11  23:59	&N02W18\\
\hline  
\end{tabular}
\end{table}

\begin{table} [ht!]
\centering
\begin{tabular}{|c|c|cc|cc|}
\cline{3-6}
\multicolumn{2}{c|}{}          &              \multicolumn{4}{c|}{\cellcolor{Gray}{2004}} \\
\hline   
10564&	X&	 	 23/02   00:00	&	N13E26	&	26/02  23:59	&N14W27\\
\hline   
10649&	X&	 	 14/07  00:00	&	S10E64	&	19/07  23:59	&S10E00\\
\hline   
10652&	M5$<$&		 21/07   00:00	&	N10E32	&	22/07  23:59	&N08E06\\
\hline   
10691&	X&	 	 29/10   14:00	&	N15W02	&	30/10  23:59	&N14W25\\
\hline
10696&	X&	 	 04/11   00:00	&	N09E32	&	10/11  23:59	&N08W62\\
\hline        
10715&	X&		 30/12   00:00	&	N04E61	&	31/12  23:59	&N04E34\\
\hline                   
\end{tabular}
\end{table}

\begin{table} [ht!]
\centering
\begin{tabular}{|c|c|cc|cc|}
\cline{3-6}
\multicolumn{2}{c|}{}          &              \multicolumn{4}{c|}{\cellcolor{Gray}{2005}} \\
\hline   
10720&	X&	 	 12/01   00:00	&	N13E52	&	20/01  23:59	&N14W70\\
\hline   
10759&	M5$<$&	 	 11/05   00:00	&	N12E50	&	13/05  23:59	&N12E06\\
\hline  
\end{tabular}
\end{table}

\begin{table} [ht!]
\centering
\begin{tabular}{|c|c|cc|cc|cc|}
\cline{3-6}
\multicolumn{2}{c|}{}          &              \multicolumn{4}{c|}{\cellcolor{Gray}{2006}} \\
\hline   
10875&	M5$<$&	 	 25/04   00:00	& S10E62	&	27/04  23:59	&S11E20\\
\hline   
10930&	X&	 	 11/12   00:00	& S05E06 &	15/12  23:59	&S06W59\\
\hline  
\end{tabular}
\end{table}

\begin{table} [ht!]
\centering
\begin{tabular}{|c|c|cc|cc|}
\cline{3-6}
\multicolumn{2}{c|}{}          &              \multicolumn{4}{c|}{\cellcolor{Gray}{2010}} \\
\hline   
11045&	M5$<$&	 	 06/02   04:00	& N24E20	&	08/02  23:59	&	N23W17\\
\hline   
11046&	M5$<$&	 	 10/02   00:00	& N24E42&	12/02  23:59	&	N24E00\\
\hline
11066&	B &	 	 03/05   00:00	& S27E16&	03/05  23:59	&	S27E04\\
\hline
11069&	M1$<$ &	 05/05   00:00	& N40W20&	07/05  23:59	&	N40W63\\
\hline   
11078&	B&	 	 08/06 08:00	&S21W40&	09/06 23:59	&	S21W61\\
\hline   
11081&	C&	 	 12/07 00:00	& N23W45&	13/07 10:00	&	N23W66\\
\hline   
11092&	C&		 31/07   18:00	& N13E50&	01/08  23:59	&	N13E20\\
\hline   
11099&	C&	 	 13/08   11:00	&N19W42&	14/08  23:59	& N19W60\\ 
\hline   
11109&	C&	 	 13/08   11:00	&N19W42&	14/08  23:59	& N19W60\\ 
\hline   
11117&	C&	 	 24/10   00:00	&S22E24&	11/10  12:00	&	S22W70\\ 
\hline   
11123&	C&	     11/10   12:00	&N20E23&	11/10  12:00	&	N20W16\\ 
\hline   
11130&	C&	 	 30/11   00:00	& N13W54&	02/12  23:59	&	N13W54\\ 
\hline  
\end{tabular}
\end{table}

\begin{table} [ht!]
\centering
\begin{tabular}{|c|c|cc|cc|}
\cline{3-6}
\multicolumn{2}{c|}{}          &              \multicolumn{4}{c|}{\cellcolor{Gray}{2011}} \\
\hline   
11142&	C&	 	 03/01 18:00	&S14E11	&	03/01 23:35	&	S14E08\\
\hline   
11158&	X&	 	 12/02   00:00	&S19E25	&	15/02  23:59	&	S21W27\\
\hline   
11164&	M1$<$&	 	 05/03   00:00	& N2319	&	07/03  23:59	&	N23W58\\
\hline   
11166&	X&	 	 07/03   00:00	&N11E27	&	11/03  17:00	&	N09W36\\
\hline   
11169&	M1$<$&	 	 12/03   00:00	&N17W11	&	15/03  23:59	&	N17W65\\
\hline   
11176&	M1$<$&	 	 24/03   00:00	&S15E56	&	25/03  23:59	&	S15E30\\
\hline   
11190&	M1$<$&	 	 14/04   08:00	&N13W05	&	17/03  23:59	&	N13W55\\
\hline   
11204&	B&	 	 09/05   00:00	&N17W43	&	11/05  03:00	&	N17W60\\
\hline   
11210&	C&	 	 09/05   00:00	&N20E20	&	10/05  23:59	&	N20W08\\
\hline   
11224&	C&	 	 28/05   00:00	&N21W15&	30/05  00:00	&	N21W55\\
\hline   
11226&	M1$<$&	 	 05/06    06:00	&S22W27&	07/06  07:00	&	S22W55\\
\hline   
11227&	C&	 	 31/05   18:00	&S20E66&	02/06  08:00	&	S20E27\\
\hline   
11236&	C&	 	 20/06    18:00	&N17E23&	21/06  23:59	&	N17W60\\
\hline   
11241&	B&		 25/06    04:00	&N20W05&	27/06 12:00	&	N20W40\\
\hline   
11244&	C&	 	 03/07    00:00	&N16W25&	03/07 23:59	&	N16W40\\
\hline   
11249&	C&	 	 09/07    19:00	&S19E01&	11/07 12:00	&	S19W22\\
\hline   
11260&	M1$<$&	 26/07    16:00	&N19E50&	28/07 19:00	&	N19E20\\
\hline   
11261&	M5$<$&	29/07    00:00	&N16E48&	05/08 23:59	&	N16W64\\
\hline   
11281&	C&	 	 31/08    19:00	&S20E30&	04/09 23:59	&	S20W12\\
\hline   
11283&	X&	 	 04/09     00:00	&N13E22&	10/09 22:00	&	N13W70\\
\hline   
11363&	C&	 	 02/12     17:00	&S21E35&	06/12 20:00	&	S21W20\\
\hline   
11387&	M1$<$&	 	 25/12     04:00	&S21E36&	27/12 23:59	&	S21W57\\
\hline
\end{tabular}
\end{table}

\begin{table} [ht!]
\centering
\begin{tabular}{|c|c|cc|cc|}
\cline{3-4}
\multicolumn{2}{c|}{}          &              \multicolumn{4}{c|}{\cellcolor{Gray}{2012}} \\
\hline   
11402&	M5$<$&	 	 20/01   00:00	& N24E16	& 23/01  23:59	&N30W24\\
\hline   
11429&	X&		 04/03   00:00	& N17E67	& 11/03  12:00	&N17W31\\
\hline   
11430&	X&		 05/03   00:00	& N20E38	& 07/03  23:59	&N17E12\\
\hline  
11455&	B&		 13/04   00:00	& N06E06	& 14/04  08:00	&N14W24\\
\hline  
11465&	C&		 21/04   00:00	& S18E40& 27/04  23:59	&S18W55\\
\hline  
11476&	M5$<$&		 07/05   00:00	& N10E60& 13/05  23:59	&N10W37\\
\hline  
11490&	B&	 	 28/05   00:00	& S12E17& 28/05  15:00	&S12E08	\\

\hline  
11494&	M1$<$&		 05/06   18:00	& S17E20& 08/06  23:59	&S17W32\\
\hline  
11504&	M1$<$&	 	 12/06   00:00	& S17E40& 14/06  19:00	&S17E03\\
\hline  
11512&	C&	         26/06   00:00	& S16E40& 29/06  18:00	&S16W14\\
\hline  
11515&	X&		 01/07   00:00	& S17E30& 07/07  23:59	&S17W65\\
\hline  
11520&	X&	 	 09/07   00:00	& S16E46& 13/07  00:00	&S17W60\\
\hline  
11542&	 C&	 08/08  09:00	& S14E63& 12/08  08:00	&S14W05\\

\hline  
11553&	 C&	 	 30/08  09:00	& S20W16& 02/09  23:59	&S20W68\\
\hline  
11613&	 M5$<$&	 		 12/11  00:00	& 	S22E57& 13/11  23:59&S22E31\\
\hline  
11618&	 M1$<$&	 	 19/11  00:00	& S12E40& 26/11  23:59	&N06W66\\
\hline   
\end{tabular}
\end{table}

\begin{table} [ht!]
\centering
\begin{tabular}{|c|c|cc|cc|}
\cline{3-6}
\multicolumn{2}{c|}{}          &              \multicolumn{4}{c|}{\cellcolor{Gray}{2013}} \\

\hline   
 11719&	 M5$<$&	 	 08/04  17:00	& N10E53& 11/04  23:59	&N11W14\\
 \hline   
 11776&	 C&	 	 18/06  08:00	& N11E11& 19/06  18:00	&N11W07\\
  \hline   
 11818&	 M1$<$&	 	 16/08  04:00	& S07W04& 17/08  23:59	&S07W35\\
\hline   
 11865&	 M1$<$&	 	 09/10  16:00	& S22E60& 15/10  15:00	&S22W22\\
  \hline   
 11875&	 X&	 	 19/10  00:00	& N06E16& 28/10  03:00	&N06W68\\
 \hline   
 11877&	 M5$<$&	 	 23/10  00:00	& S12W08& 24/10  10:00	&S12W15\\
  \hline   
 11884&	 M5$<$&	 	 29/10  04:00	& S12E52& 03/11  17:00	&S12W21\\
\hline   
 11890&	 X&	 	 04/11  00:00	& S11E63& 12/11  23:59	&S11W58\\
\hline   
 11936&	 M5$<$&	 	 30/12  00:00	& S16W09& 02/01  23:59	&S16W67\\
\hline   
\end{tabular}
\end{table}

\begin{table} [ht!]
\centering
\begin{tabular}{|c|c|cc|cc|}
\cline{3-6}
\multicolumn{2}{c|}{}          &              \multicolumn{4}{c|}{\cellcolor{Gray}{2014}} \\
\hline   
 11944&	 X &	    05/01   00:00	& S09E40&  10/01  09:00	&S09W33\\
 \hline   
 11966&	 M5$<$&	 10/03   10:00	& N14W41&  12/03    22:00 &N14W72\\
\hline   
 11967&	 M5$<$&	  31/01   00:00	& S12E44&  08/02    23:59 &S12W63\\
 \hline   
 12017&	 X &  28/03   00:00	& N10W08&  30/03  23:59	&N10W50\\
 \hline   
 12036&	 M5$<$ &15/04    00:00   &	S17E13&  18/04    23:59   &S17W40\\
  \hline   
 12146&	 M1$<$&	 22/08    19:00	& N09W01&  25/08  23:59	&N09W43\\
  \hline   
 12158&	 X &		   07/09    15:00	& N15E54&  11/09  23:59	&N15W12\\
 \hline   
 12192&	 X &	 	   18/10    09:00 & S13E70&  27/10   15:00 &S13W52\\
\hline   
 12205&	 X &	 	    05/11    12:00 & N15E66&   12 /11  23:59 &N15W35\\
\hline   
 12241&	 M5$<$&		    18/12    00:00 & S09E20&   21/12   23:59 &S09W34\\
 \hline 
 12246&	 M5$<$&	    18/12    00:00 & S09E20&   21/12   23:59 &S09W34\\
\hline   
12242&	 X &	 	    15/12    15:00 & S17W33&   20/12   23:59 &S17W33\\
 \hline   
\end{tabular}
\end{table}

\end{document}